\documentclass[12pt]{iopart}
\usepackage{graphicx}
\usepackage{url}
\usepackage{iopams}
\usepackage{booktabs}
\usepackage[T1]{fontenc}
\usepackage[percent]{overpic}
\newcommand{\panel}[1]{\textsf{\small (#1)}}
\graphicspath{{figures/}}

\begin{document}

\title[Emergent topological structure in brain-organoid activity]{Emergent topological structure in spontaneous brain-organoid activity}

\author{Eve Bodnia$^{1,2,\dagger}$, Margaux Basart$^{3,4,\dagger}$, Sofie Hai$^1$, Lenzie Ford$^1$, Nina Miolane$^5$, Kenneth S. Kosik$^{1,6}$, Dirk Bouwmeester$^{2,7}$, Lincoln D. Carr$^{3,4,8,9,\ast}$}

\address{$^1$ Department of Molecular, Cellular, and Developmental Biology, University of California, Santa Barbara, CA 93106, USA}
\address{$^2$ Department of Physics, University of California, Santa Barbara, CA 93106, USA}
\address{$^3$ Department of Physics, Colorado School of Mines, Golden, CO 80401, USA}
\address{$^4$ Quantum Engineering Program, Colorado School of Mines, Golden, CO 80401, USA}
\address{$^5$ Department of Electrical and Computer Engineering, University of California, Santa Barbara, CA 93106, USA}
\address{$^6$ Neuroscience Research Institute, University of California, Santa Barbara, CA 93106, USA}
\address{$^7$ Huygens-Kamerlingh Onnes Laboratory, Leiden University, P.O.\ Box 9504, 2300 RA Leiden, the Netherlands}
\address{$^8$ Department of Applied Mathematics and Statistics, Colorado School of Mines, Golden, CO 80401, USA}
\address{$^9$ Institute of Science and Technology Austria (ISTA), Am Campus 1, 3400 Klosterneuburg, Austria}

\vspace*{5pt}
\address{$^{\dagger}$These authors contributed equally to this work.}
\address{$^{\ast}$E-mail: \mailto{lcarr@mines.edu}}

\begin{abstract}
Neural activity is widely held to organize on low-dimensional structure embedded
in a high-dimensional state space. Persistent homology reads such structure
directly from the pattern of pairwise correlations, without assuming in advance
which variables are relevant. We apply persistent homology to
microelectrode-array (MEA) recordings of spontaneous activity from human (Lancaster) and mouse (Pa\c{s}ca)
cortical organoids, spanning $25$--$197$ sorted units, and ask
whether topological data analysis resolves structure at the node counts
that neural recordings actually deliver. Building weighted networks in correlation space
and characterizing them by Vietoris--Rips filtration, we find that the
first homology ($H_1$, loops) rises significantly above a rate- and
population-preserving null in $14$ of $18$ datasets, all of which remain significant
after correction for testing $18$ datasets. Random removal of
neural units leaves the loops largely intact. Removing the units that form the loops
disrupts them more than removing equally connected units does. Topological richness grows with network size, and second
homology ($H_2$) exceeds the null only in the larger networks. These
results show that persistent homology resolves structured topology in neural recordings
at the scale experiments actually deliver.
\end{abstract}

\maketitle

% =====================================================================
\section{Introduction}
\label{sec:intro}

A recurring idea in systems neuroscience is that meaningful neural activity
does not fill its high-dimensional state space but concentrates on a
lower-dimensional manifold, the neural manifold hypothesis~\cite{Singh2008,gardner_2022,Acosta2022QuantifyingLE}.
Where that manifold is curved or multiply connected, linear tools such as principal
component analysis report the manifold's dimension but miss its shape. Topological data analysis
(TDA), and persistent homology in particular, recovers that shape, in the form of
topological structure: the number and persistence of connected components, loops, and
voids, counted respectively by the Betti numbers $\beta_0,\beta_1,\beta_2$. These are
computed directly from a matrix of pairwise relationships and are invariant to any
monotone rescaling of that matrix~\cite{Giusti2015,Sizemore2019}.

Applied to neural data, persistent homology has revealed the circular and toroidal
geometry of spatial coding~\cite{gardner_2022}, intrinsic structure in
hippocampal correlations independent of receptive-field models~\cite{Giusti2015},
cavities and cliques in structural connectomes and reconstructed
microcircuits~\cite{Reimann2017,Sizemore2018}, and homological organization of
functional networks~\cite{Petri2014,Sizemore2019}. In parallel, physicists studying
complex systems have adopted higher-order interactions and simplicial
structure as a common language~\cite{Battiston2020,Battiston2021,BassettSporns2017}.
The clique complex of a correlation matrix is built from pairwise relations, so its higher-dimensional homology describes
how pairwise co-firing closes into loops and voids, not interactions among three or
more units~\cite{Battiston2020}.

Brain organoids, self-organizing three-dimensional cultures of human or mouse
cortical tissue~\cite{Lancaster2013,Birey2017}, are an increasingly important
model system. High-density MEAs can now record the spontaneous spiking of hundreds
of units in a single organoid at once~\cite{Sharf2022Functional}. Graph-theoretic network analysis
has already been brought to organoid MEA data~\cite{MEANAP2024}, characterizing degree,
clustering, and modularity. Persistent homology characterizes the global organization
of that same correlation graph, the loops and voids that local measures of this kind
miss, and has not previously been applied
to the spontaneous activity of brain organoids.

Real neural recordings, in organoids, in slices, and in vivo, rarely resolve more
than a few hundred units at once, far short of the $10^4$--$10^6$ in a full circuit.
MEA is our starting point because its sub-millisecond sampling captures exact spike
times, and the correlations we analyze are built on that timing; optical methods
are one to two orders of magnitude slower and blur it. Across eighteen organoid datasets, persistent homology
finds loop topology over and above what a rate- and population-preserving null
produces, at the unit counts experiments actually deliver.

% =====================================================================
\section{Organoid MEA data}
\label{sec:data}

We analyze spontaneous extracellular recordings from two organoid protocols: human
cerebral organoids (Lancaster protocol; datasets labeled O$n$) and mouse cortical
organoids (Pa\c{s}ca protocol; datasets labeled MO$n$). The protocols differ in ways
that shape the networks: Lancaster organoids self-organize without regional
patterning and develop numerous rosettes spanning several presumptive brain regions,
whereas Pa\c{s}ca organoids are directed toward a cortical fate and carry one or few
rosettes~\cite{Lancaster2013,Birey2017}. Activity was
recorded on high-density CMOS MEAs at $20$\,kHz for $3$\,min (O1--O4), $10$--$12.7$\,min
(MO1--MO12), or $30$\,min (O5, O6),
resolving spike times to sub-millisecond precision, and spike-sorted with
Kilosort2~\cite{Pachitariu2024}. The sorted single units that passed quality control and
the duplicate check below form the nodes of all networks. The resulting datasets span
$N=25$ to $N=197$ units (Table~\ref{tab:datasets}).

\begin{figure}[t]
\centering
\IfFileExists{figures/fig_isi.pdf}{%
  \includegraphics[width=\textwidth]{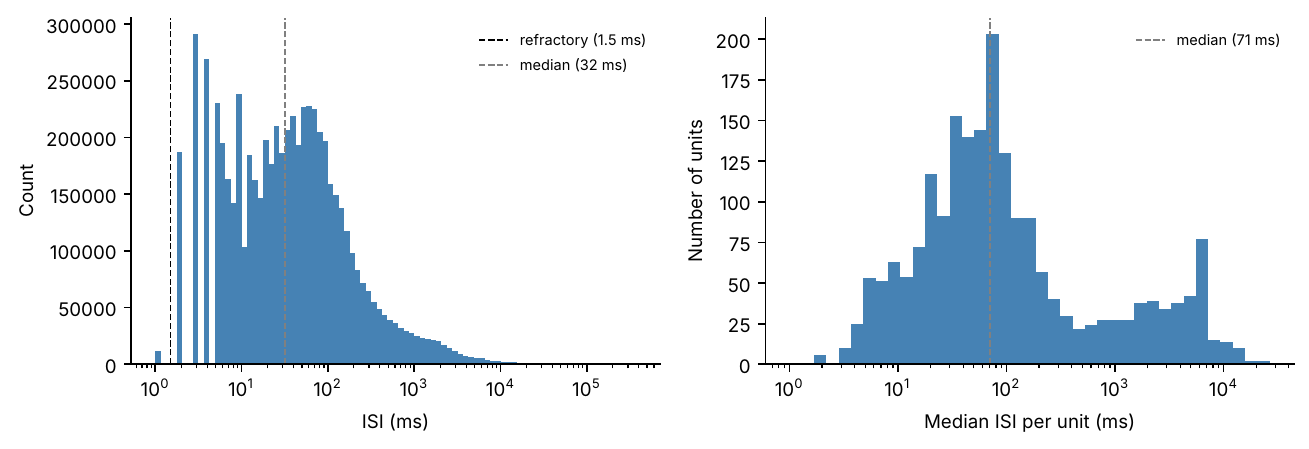}%
}{%
  \fbox{\begin{minipage}[c][3.2cm][c]{0.86\textwidth}
  \centering
  [\,interspike interval statistics: pooled ISI distribution with refractory
  structure, and distribution of per-unit median ISI\,]
  \end{minipage}}%
}
\caption{Spike-train statistics across all datasets. (Left) Pooled interspike interval
(ISI) distribution over all units and recordings; counts fall to near zero below the
$1.5$\,ms refractory period, confirming well-isolated single units, and the median
pooled ISI is $32$\,ms. (Right) Distribution of per-unit median ISI, whose median
across units is $71$\,ms.
Both are consistent with bursting spontaneous activity.}
\label{fig:isi}
\end{figure}

We compute correlations within short lag windows of $0$, $10$, and $20$\,ms
(Sec.~\ref{sec:corr}), which probe co-firing at successively longer latencies to
allow for conduction and synaptic delays between units; results reported here use the
$0$\,ms window unless noted. Spike trains are smoothed with a $50$\,ms Gaussian kernel
before correlation. This kernel width sets the timescale on which co-firing is
measured~\cite{Schreiber2003}, and $50$\,ms matches the tens-of-millisecond scale of
synaptic interaction among neurons. The
activity is strongly bursting, which the null model of Sec.~\ref{sec:corr} controls
for. The sorted units show clear refractory structure, with $0.19\%$ of intervals
shorter than the $1.5$\,ms refractory period, and a pooled median interspike interval
of $32$\,ms (Fig.~\ref{fig:isi}).

Because one neuron can be sorted into several units on a high-density array, we flag
a pair as duplicates when more than half of the spikes of the less active unit fall
within $\pm1$\,ms of a spike of the other, with at least $10$ such coincidences and a
zero-lag count more than ten times that expected from lags of $4$ to $13$\,ms. We keep
the most active unit of each flagged group, which removes $361$ of $2{,}419$ units. A
lag-aware variant, which also catches copies offset by up to $5$\,ms and ignores units
with fewer than $100$ spikes, removes $228$ units. With it the data exceed all $19$ surrogates in the datasets that
are significant in the main test, and in MO11 as well (Table~\ref{tab:stats}).

\begin{table}[t]
\centering
\small
\caption{Datasets, unit counts before ($N_0$) and after ($N$) duplicate removal, and
recording duration $T$. Human: Lancaster protocol (O$n$); mouse cortical: Pa\c{s}ca
protocol (MO$n$).}
\label{tab:datasets}
\begin{tabular}{lrrr@{\hskip 2.5em}lrrr}
\toprule
Dataset & $N_0$ & $N$ & $T$ (min) & Dataset & $N_0$ & $N$ & $T$ (min)\\
\midrule
O1 & 26  & 25  & 3  & MO5  & 199 & 188 & 10.1 \\
O2 & 131 & 120 & 3  & MO6  & 170 & 146 & 10.4 \\
O3 & 40  & 36  & 3  & MO7  & 234 & 197 & 10.0 \\
O4 & 123 & 122 & 3  & MO8  & 194 & 168 & 10.1 \\
O5 & 151 & 146 & 30 & MO9  & 86  & 78  & 11.0 \\
O6 & 119 & 85  & 30 & MO10 & 157 & 125 & 10.5 \\
MO1 & 215 & 179 & 10.0 & MO11 & 82 & 28 & 12.7 \\
MO2 & 211 & 165 & 10.0 & MO12 & 48 & 44 & 10.0 \\
MO3 & 130 & 113 & 10.1 &  &  &  & \\
MO4 & 103 & 93  & 10.1 &  &  &  & \\
\bottomrule
\end{tabular}
\end{table}

\begin{figure}[t]
\centering
\includegraphics[width=\textwidth]{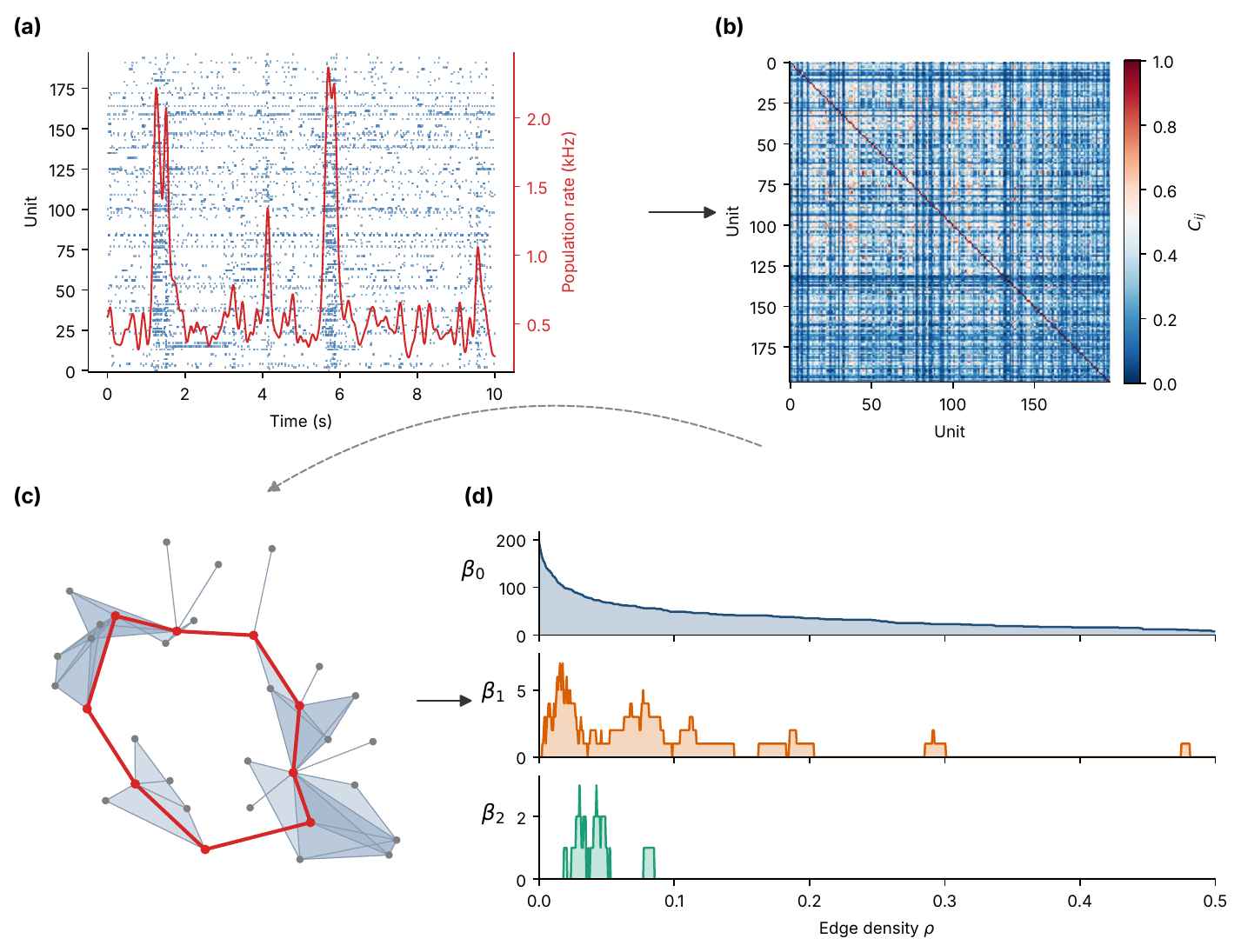}
\caption{Analysis pipeline, shown for MO7. (a) Spike-sorted units and the population
rate. (b) The pairwise correlation matrix $C_{ij}$. (c) The matrix defines a
Vietoris--Rips filtration on $d_{ij}=1-C_{ij}$, from which persistence reads loops and
voids. Shown is a neighborhood of MO7's most persistent $H_1$ loop at the scale where it forms, drawn as units (nodes), 
co-firing edges, and filled 2-simplices (shaded), with the
loop highlighted in red enclosing the hole it bounds. Node positions come from
multidimensional scaling of the correlation distances, not electrode coordinates.
(d) Persistent homology is read out as Betti curves $\beta_0$,
$\beta_1$, $\beta_2$ against edge density $\rho$.}
\label{fig:pipeline}
\end{figure}

% =====================================================================
\section{Pairwise time correlations and the null model}
\label{sec:corr}

For each pair of units $(a,b)$ we form a correlation $C(a,b)\in[0,1]$ from the
overlap of their Gaussian-smoothed spike trains,
\begin{equation}
C(a,b)=\max_{j\in J}\;
\frac{\sum_i a_i\,b_{i+j}}{\sqrt{\sum_i a_i^2}\,\sqrt{\sum_i b_i^2}},
\label{eq:corr}
\end{equation}
the largest normalized overlap of the two smoothed trains over lags $j$ in a window
$J=[-w,w]$, with $C=1$ for identical activity. At zero lag this is the Gaussian-smoothed
spike-train correlation of Schreiber et al.~\cite{Schreiber2003}. Built from co-activation, $C$ is Hebbian in
character, registering units that fire together and underweighting anti-correlated,
inhibitory-like coupling.

For $w>0$ the maximum over lags can only raise $C$, most for pairs whose co-firing is
offset in time, and data and surrogates pass through the same operation. We repeated
the $H_1$ test at $w=10$ and $20$\,ms against $20$ surrogates per dataset. The data exceed all of them in $13$ of the $14$ datasets significant at zero
lag, and integrated $\beta_1$ changes by a median of $5$--$6\%$ in the $13$ datasets
where it exceeds $0.04$. Because $J$ is symmetric about
zero, $C$ is symmetric, as the undirected filtration below requires. Each unit is then
a point in a \emph{correlation space} where proximity means functional similarity, not
anatomical distance. A correlation links units that co-fire and need not mark a direct
synaptic connection: the network is functional, not structural. We test directly
whether the resulting topology is intrinsic or inherited from the electrode layout, in
the five datasets that retain electrode coordinates (Sec.~\ref{sec:r5}).

\begin{figure}[t]
\centering
\includegraphics[width=0.6\textwidth]{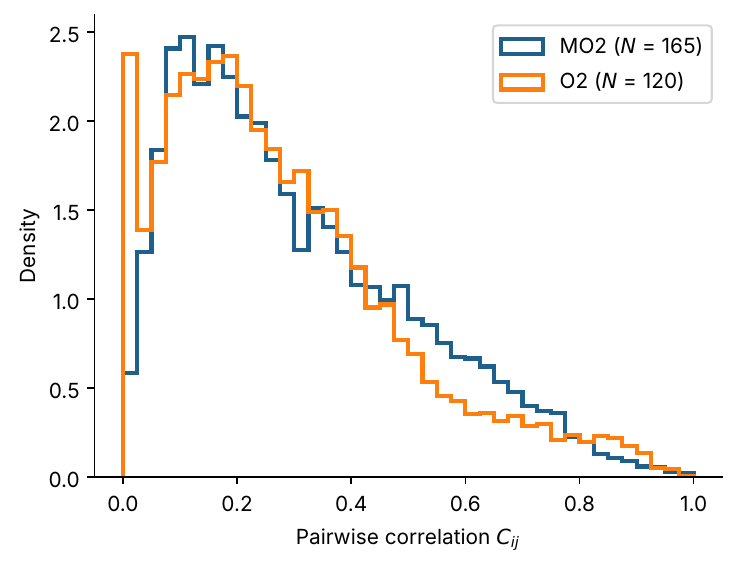}
\caption{Pairwise correlation distributions for a representative human (O2) and
mouse (MO2) dataset: both are right-skewed, with most pairs weakly to
moderately correlated and a tail of strongly co-active pairs.}
\label{fig:corr}
\end{figure}

In the two datasets of Fig.~\ref{fig:corr} the correlation distributions are
right-skewed. Most unit pairs are weakly to moderately correlated, with a peak
near $C\approx0.1$--$0.2$ and a long tail of strongly co-active pairs, and about a tenth
of all pairs exceed $C=0.6$. This tail is a signature of the bursting dynamics in which
sub-populations co-activate, reported before in cortical
organoids~\cite{Sharf2022Functional,Osaki2024,Wilson2022}, and it motivates a topological treatment: the
strong-correlation backbone those pairs form defines a sparse graph whose clique
structure is the object of interest.

To test whether topological structure exceeds what
firing rate and population bursting alone produce, we randomize the binary spike
raster under a constraint: the \emph{raster-marginals}
model~\cite{Okun2012}. Each randomization holds fixed both margins of the
units$\times$time-bins raster ($1$\,ms bins), every unit's total spike count and every
time bin's total population activity, while destroying the specific co-firing of
particular units, by repeatedly
selecting $2\times2$ submatrices of the form
$\left(\begin{array}{@{}cc@{}}1&0\\0&1\end{array}\right)$ or
$\left(\begin{array}{@{}cc@{}}0&1\\1&0\end{array}\right)$ and interchanging the two
patterns. Such
interchanges connect all $0/1$ matrices with given margins~\cite{Ryser1957}, but a
chain of finite length must also mix~\cite{Miklos2004}. We attempt $10$ interchanges
per spike, and integrated $\beta_1$ and the correlations of the surrogates converge by $2$ attempts per spike (Fig.~\ref{fig:convergence}).
Correlations are recomputed on each randomized raster, and the entire topological
analysis is repeated on the surrogate ensemble.
Because the surrogates preserve each unit's firing rate and the population activity
in each time bin, the two properties that dominate organoid spiking~\cite{Okun2012},
any topology that exceeds the null reflects co-firing among specific units that these
marginals leave undetermined. In the ten- and thirty-minute recordings other than MO11, at most $5$ of $100$
surrogates contain any loop. Their correlations are close to a function of
single-unit properties, since a fit $\log C_{ij}=a_i+a_j$ reproduces their rank order
with Spearman coefficient $0.998$ in MO7, against $0.84$ for the data, and a matrix of
exactly that form has no loops at any edge density. The
two halves of each recording give correlation matrices that agree with median Pearson
$r=0.97$, and both halves contain loops in every dataset with significant $H_1$. A fully random
(Erd\H{o}s--R\'enyi) network serves only as a structural reference, not a
significance test.

% =====================================================================
\section{Topological methods}
\label{sec:topo}

\begin{figure}[t]
\centering
\includegraphics[width=\textwidth]{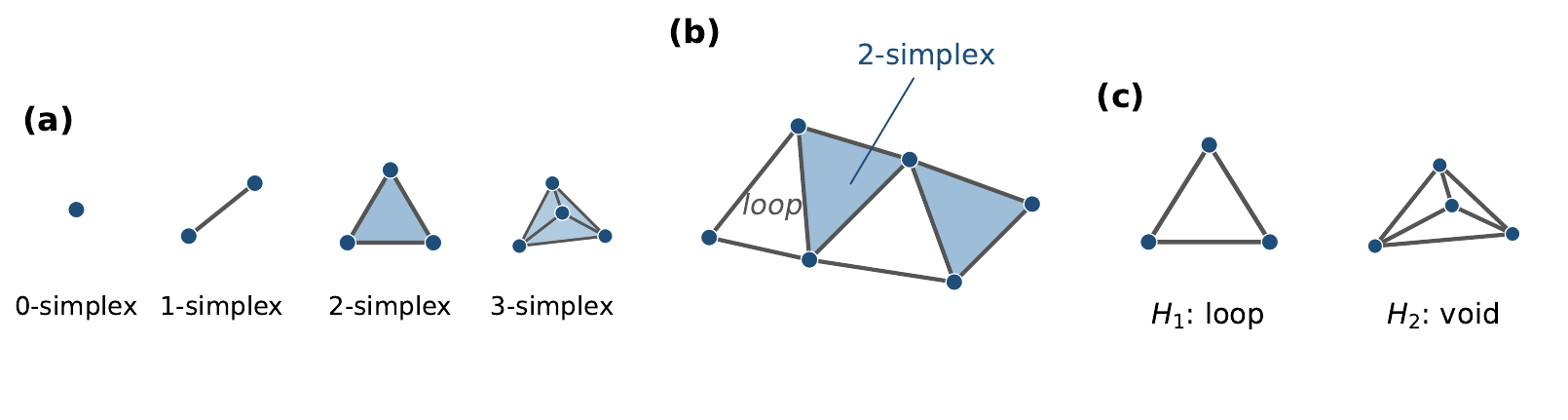}
\caption{Building blocks of persistent homology. A filled (blue) region is a simplex
present in the complex; a region bounded by edges or faces but left unfilled is a hole.
(a) Simplices of dimension $0$ to $3$: a point, an edge, a filled triangle, and a
filled tetrahedron. (b) A complex glued from such simplices: the blue triangles are
$2$-simplices, while the three edges at lower left bound an open triangle that no
$2$-simplex fills, a loop. (c) The holes the method counts: a loop that no simplex
fills is a first-homology ($H_1$) feature, and an enclosed cavity is a second-homology
($H_2$) feature.}
\label{fig:simplex}
\end{figure}

Persistent homology represents the correlation network as a growing family of
simplicial complexes and counts the holes in them (Fig.~\ref{fig:simplex}). The
building blocks are simplices: a point, an edge, a filled triangle, and a filled
tetrahedron are the simplices of dimension $0$ through $3$, and a simplicial complex
is a collection of them joined along shared faces.

\begin{figure}[t]
\centering
\includegraphics[width=\textwidth]{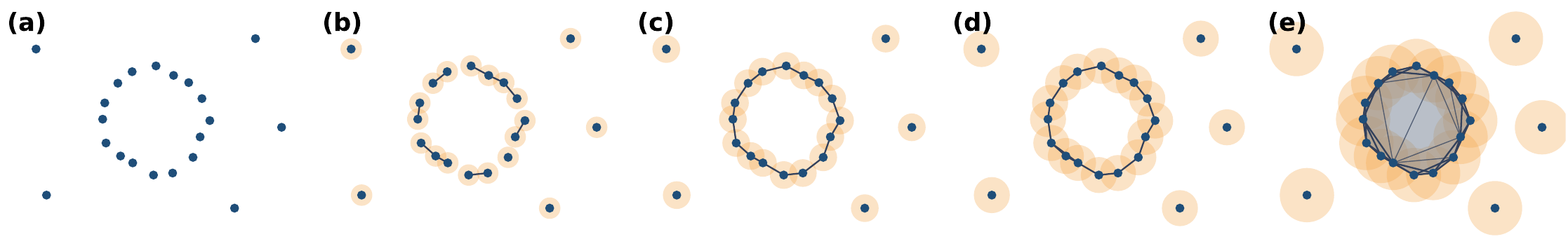}
\caption{Vietoris--Rips filtration, drawn in two dimensions for intuition; the real
filtration runs in the high-dimensional correlation space. Each unit is a point (a),
and as the scale $\varepsilon$ grows, every unit carries a disk of radius
$\varepsilon/2$; two units join by an edge when their disks overlap
($d_{ij}\le\varepsilon$), and mutually joined sets fill in as simplices. A loop forms
once the ring of edges closes (c), survives as $\varepsilon$ increases (d), and dies
when the interior fills (e). The range of $\varepsilon$ over which it survives is its
persistence, which separates a real feature from noise.}
\label{fig:filtration}
\end{figure}

Persistence is computed on the full correlation matrix with
\texttt{Ripser}~\cite{ctralie2018ripser} (version 0.6.15, homology up to dimension $2$,
no distance threshold). We convert each correlation matrix to a
dissimilarity $d_{ij}=1-C_{ij}$ and build
the Vietoris--Rips (clique) filtration: as a scale $\varepsilon$ increases from $0$
to $1$, an edge
appears between units $i$ and $j$ once $d_{ij}\le\varepsilon$, and any set of mutually
connected units is filled in as a simplex (Fig.~\ref{fig:filtration}). The $k$th Betti number
$\beta_k(\varepsilon)$ counts the independent $k$-dimensional homological features
present at scale $\varepsilon$: $\beta_0$ connected components, $\beta_1$ loops,
$\beta_2$ enclosed voids. As $\varepsilon$ grows, each feature is born at the scale
$\varepsilon_{\rm birth}$ where its cycle first closes and dies at $\varepsilon_{\rm death}$
where that cycle is filled in.

Rather than fix a single scale, we report \emph{density-indexed} Betti curves, the
clique-topology construction introduced for neural correlation matrices by Giusti
et al.~\cite{Giusti2015}.
The edge density $\rho\in[0,1]$ is the fraction of the ${N \choose 2}$ possible pairs
present at a given scale; it is a monotone reparametrization of $\varepsilon$. Writing
$F$ for the empirical cumulative distribution function of the pairwise correlations,
$\rho = 1-F(\theta)$ at correlation threshold $\theta=1-\varepsilon$, the fraction of
pairs whose correlation exceeds $\theta$. Plotting $\beta_k(\rho)$ and reporting the
integrated Betti value $\int \beta_k(\rho)\,d\rho$ has two virtues: it does not depend
on the particular correlation scale, and it places data and surrogate on a common,
rank-matched axis, so the null comparison is made at equal sparsity even though the surrogates'
correlation values differ in scale. We use the undirected construction throughout;
directed (flag) complexes are an extension to directed data. A separate threshold at the
$50$th density percentile is used only for the supporting graph-theoretic measures.
Each loop or void is a point $(\varepsilon_{\rm birth},\varepsilon_{\rm death})$ in a
persistence diagram. The bottleneck distance between two diagrams is the smallest
$\delta$ for which every point of one can be matched to a point of the other, or to
the diagonal, within $\delta$. It measures how much the topology changes between two
conditions, and we use it in Sec.~\ref{sec:r2}.

A Vietoris--Rips complex can have homology in arbitrarily high degree even for points in
the plane~\cite{Adamaszek2017}, so the homological dimension of a feature does not reflect the
spatial dimension of the recording.

% =====================================================================
\section{Results}
\label{sec:results}

This section reports five results. First, organoid networks carry more loop ($H_1$)
structure than the rate- and population-preserving null produces
(Sec.~\ref{sec:r1}). Second, this structure rests on a non-redundant core of strongly
co-active units (Sec.~\ref{sec:r2}). Third, the number of homological dimensions a
network populates grows with its size (Sec.~\ref{sec:r3}). Fourth, enclosed voids
($H_2$) emerge only in the larger networks (Sec.~\ref{sec:r4}). Finally, the
loop structure reflects co-firing rather than the physical layout of the array
(Sec.~\ref{sec:r5}).

\subsection{Loop structure exceeds the rate- and population-preserving null}
\label{sec:r1}

Our primary statistic is the integrated first Betti number~\cite{Giusti2015},
$\int\beta_1(\rho)\,d\rho$, the area under the density-swept loop count. It
summarizes, in a single threshold-free number, how much loop structure a network
carries across the whole filtration. Because it is indexed by edge density rather
than by raw correlation, it can be compared directly between a recording and its
surrogates even though randomization shifts the correlation values
(Sec.~\ref{sec:topo}). For each dataset we compare this quantity to the distribution
obtained from $100$ raster-marginals surrogates. The null distributions are not normal
(Fig.~\ref{fig:qq}), so we report the rank-based $p$-value
$p=(r+1)/101$, where $r$ is the number of surrogates whose integrated $\beta_1$ is at least that of the
data~\cite{Phipson2010}, together
with $q$, the $p$-value corrected for testing $18$ datasets~\cite{Benjamini1995}.

\begin{figure}[t]
\centering
\begin{overpic}[width=\textwidth]{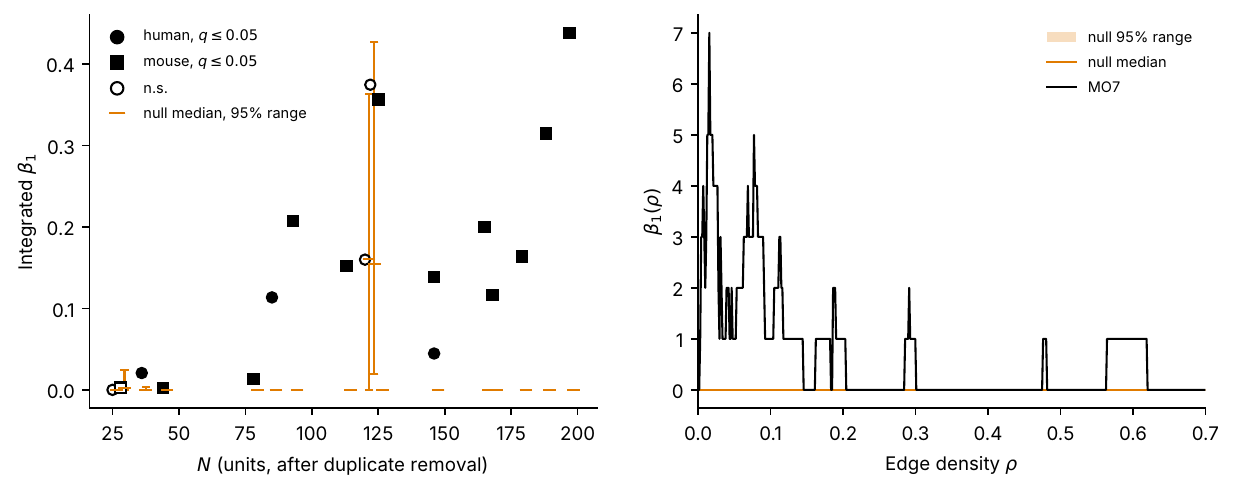}
\put(0,37.5){\panel{a}}\put(49,37.5){\panel{b}}
\end{overpic}
\caption{(a) Integrated $\beta_1$ for data (filled, $q\le0.05$; open, n.s.; circles,
human; squares, mouse) against the raster-marginals null (median and central $95\%$)
across all $18$ datasets, versus $N$. (b) $\beta_1(\rho)$ for MO7: data (black) versus
null median and central $95\%$ (orange).}
\label{fig:h1sig}
\end{figure}

Loop structure in the organoid data exceeds the surrogate null in $14$ of the $18$
datasets, all with $q\le0.025$ (Fig.~\ref{fig:h1sig}a, Table~\ref{tab:stats}). The four
that do not separate from the null are the two smallest networks (O1, $N=25$; MO11,
$28$) and two of the larger human sets (O4, $122$; O2, $120$). Size alone does not
decide the outcome. O1 has no loops, and O2 and O4 are three-minute recordings whose
surrogates also contain loops. Where the null is nearly empty, a dataset with any loops
is significant. In the mid-to-large networks integrated $\beta_1$ ranges from
$0.045$ to $0.44$.

Most loops occur at low edge density. In MO7, $\beta_1(\rho)$ peaks near $\rho=0.016$ and at most three loops
remain beyond $\rho=0.1$ (Fig.~\ref{fig:h1sig}b). As weaker correlations enter the
filtration, the loops are filled in by the triangles that high clustering supplies. A
few loops persist to $\rho\approx0.6$. The topology is therefore mainly a property of the
strong-correlation backbone: the loops are carried by the strongly co-active units, the
``choristers'' of the population rather than the weakly coupled
soloists~\cite{Okun2015}. Because the surrogates hold each unit's firing rate and each
time bin's population activity fixed, the loop structure that exceeds them cannot be a
by-product of rate or of the population-wide bursts that dominate organoid activity. It
reflects co-firing structure beyond the reach of those first-order properties.

\subsection{The loops occupy a non-redundant core}
\label{sec:r2}

If the loop structure were diffuse, supported redundantly by many interchangeable
units, it would degrade gracefully as units are removed. If instead it rests on a
specific set of carrying units, removing those should disrupt it sharply while
removing others should barely matter. We test this by deleting $10\%$ of the units
and recomputing the topology, contrasting random deletion with deletion drawn
uniformly from the supports of the $H_1$ cocycle representatives ($14$--$44\%$ of the
units), each over $100$ draws of the same number of units.

\begin{figure}[t]
\centering
\begin{overpic}[width=\textwidth]{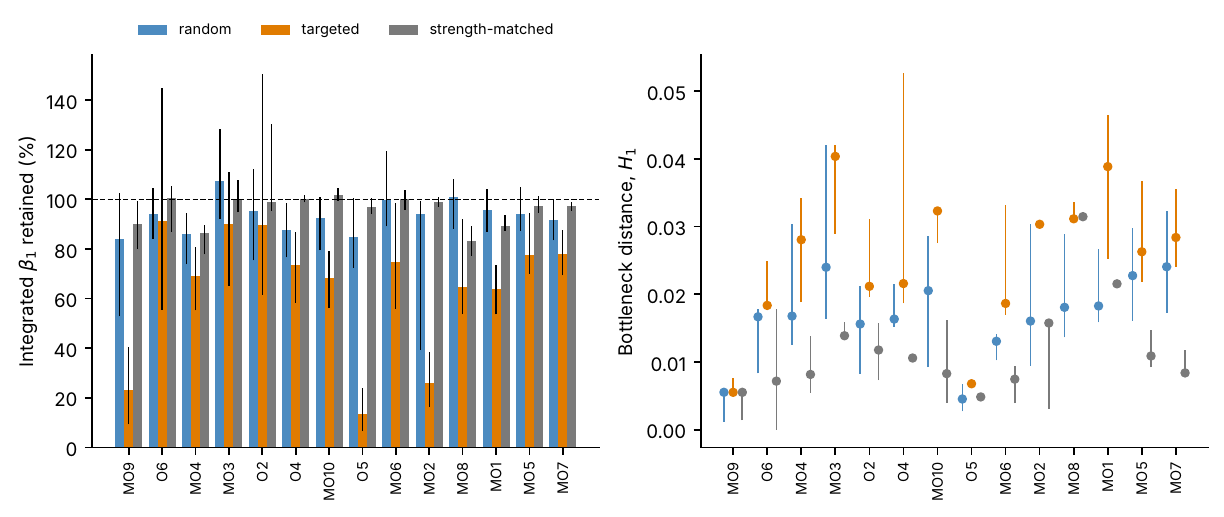}
\put(0.5,40.5){\panel{a}}\put(50.5,40.5){\panel{b}}
\end{overpic}
\caption{(a) Fraction of integrated $\beta_1$ retained under random (blue), targeted
(orange), and strength-matched (grey) removal of $10\%$ of units, median and
interquartile range over $100$ draws; dashed line: full retention. (b) Bottleneck
distance of the $H_1$ diagram under the same three conditions.}
\label{fig:removal}
\end{figure}

Random removal leaves the loop structure largely intact: across the datasets with
loop structure to test, integrated $\beta_1$ is retained at a median of $94\%$,
against $71\%$ under targeted removal (Fig.~\ref{fig:removal}a), so no small,
interchangeable subset is doing all the work. Targeted removal disrupts it far more,
but quantifying that disruption requires
care. The natural-seeming measure, the change in loop count, is in fact
misleading: deleting a densely connected hub sparsifies its neighborhood and can
\emph{open} new loops, so $\beta_1$ sometimes rises under targeted removal even as
the original loops are destroyed. We therefore measure disruption by the bottleneck
distance between the original and post-removal $H_1$ persistence diagrams, which
registers how far the loop structure has moved regardless of whether new features
appear. The mean bottleneck distance under targeted removal exceeds that under random
removal, a ratio above $1$, in all $14$ datasets with enough loop structure to
compare. We set aside the four lowest-count sets (O1, MO11, O3, MO12), whose
cocycle supports hold fewer units than are deleted. The median ratio is $1.54$, with a
range of $1.21$ to $1.66$ (Fig.~\ref{fig:removal}b, Table~\ref{tab:stats}), and a ranked test across datasets gives one-sided $p=6.1\times10^{-5}$ (Wilcoxon).
Within each dataset the $100$ targeted draws exceed the $100$ random draws
(one-sided rank-sum test, $p\le2.3\times10^{-5}$ in all $14$). Loop-supporting units are more strongly connected than the rest (median ratio of weighted degree $1.23$), so for each
targeted draw we also delete the units outside the supports closest in weighted degree.
Targeted removal moves the $H_1$ diagram more than this strength-matched removal in $13$
of the $14$ datasets, and about as much in MO8 (median ratio $2.53$, $p=6.1\times10^{-5}$). The loops are thus carried by an identifiable,
non-redundant core of units rather than spread interchangeably across the population.

\subsection{Topological richness grows with network size}
\label{sec:r3}

\begin{figure}[t]
\centering
\includegraphics[width=0.56\textwidth]{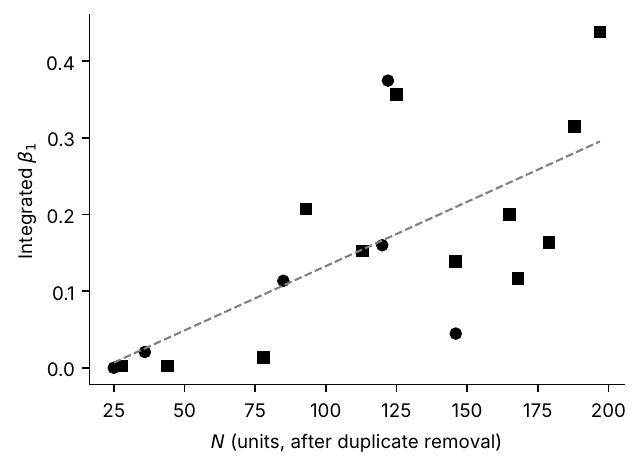}
\caption{Integrated $\beta_1$ against the number of units $N$ (circles, human; squares,
mouse; dashed line: linear fit, $r=0.68$).}
\label{fig:h1N}
\end{figure}

Richness here means how many homological dimensions a network resolves above the
null, that is, how many of $H_0$, $H_1$, $H_2$ it populates. It climbs
with $N$ because each Betti number becomes resolvable above the null only as
the network grows, with successive dimensions requiring more units. The transition is
statistical, not a sharp threshold. Connected components ($H_0$) resolve at every
size. Loops ($H_1$) need enough units to close a cycle and resolve above the null in
$14$ of $18$ datasets, not in strict order of size (Sec.~\ref{sec:r1}).
Enclosed voids ($H_2$) come online later, clearing the null only among the larger
networks (Sec.~\ref{sec:r4}). A larger network is thus more likely to resolve
each next dimension: $H_1$ appears readily and $H_2$ in the larger networks.
The count within a dimension tells the same story more weakly: integrated $\beta_1$
rises with $N$ (Pearson $r=0.68$; Fig.~\ref{fig:h1N}), but the correlation is inflated
by the small networks that resolve no loops, so the within-dimension magnitude is not a
reliable size law. Richness grows with size by adding dimensions, not loops within one.

Random
subsets of MO7 exceed their own null in $H_1$ at every size from $25$ units. In $H_2$ they
exceed it in none of $10$ subsets of $25$ units, in $7$ of $10$ subsets of $50$, and in all
subsets of $100$ and $150$
(Fig.~\ref{fig:subsample}). Within a recording voids thus need more units than loops, as
across datasets. Our networks are at the scale of the $88$ place cells of Giusti
et al.~\cite{Giusti2015} and two orders of magnitude smaller than the reconstructed
microcircuit of Reimann et al.~\cite{Reimann2017}.

\subsection{Second homology appears in the larger networks}
\label{sec:r4}

\begin{figure}[t]
\centering
\begin{overpic}[width=\textwidth]{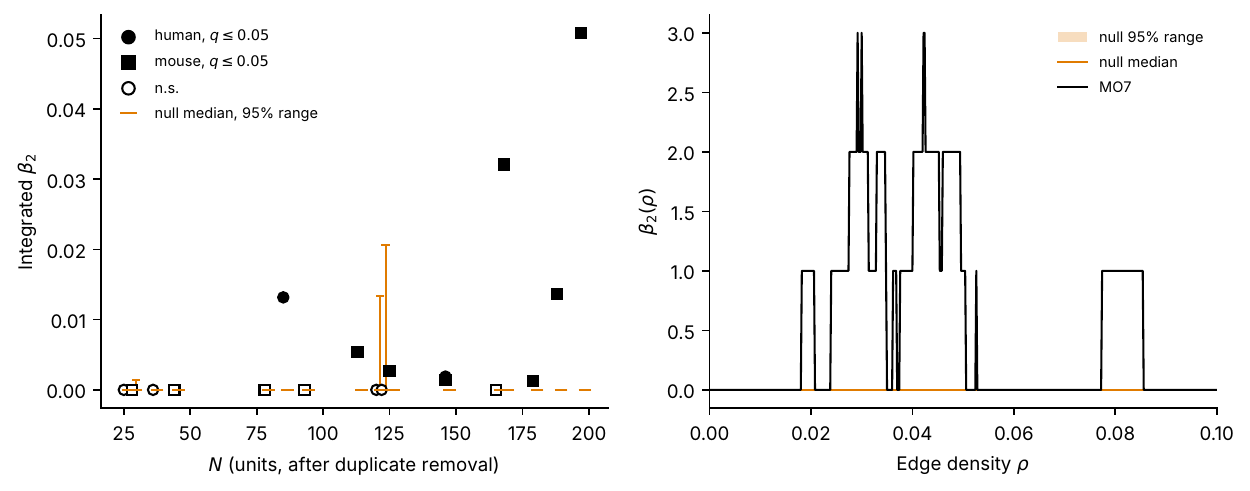}
\put(0,37){\panel{a}}\put(49,37){\panel{b}}
\end{overpic}
\caption{(a) Integrated $\beta_2$ for each dataset against unit count $N$. Filled
symbols mark datasets whose $\beta_2$ exceeds the rate- and population-preserving null
at $q\le0.05$, open symbols those that do not (circles, human; squares, mouse). Orange
marks give the null median and central $95\%$. (b) $\beta_2(\rho)$ for MO7: data (black)
and null (orange).}
\label{fig:h2}
\end{figure}

Enclosed voids ($H_2$) require both more units and richer co-firing to form than
loops. Under the same surrogate test, $H_2$ rises
significantly above the null in nine datasets (O6, MO3, MO10, O5, MO6, MO8, MO1, MO5,
MO7), each with $N\ge85$ and $q=0.020$, and most strongly in the largest
(Fig.~\ref{fig:h2}a; MO7 in Fig.~\ref{fig:h2}b). No dataset below $N=85$ shows significant $H_2$, but neither do four
of the thirteen at or above it (MO4, O2, O4, MO2), so this size is necessary in this
corpus but not sufficient. The voids that drive it are few and low-persistence, 
shorter-lived than the longest loops in the same data, so the excess is significant but not
yet a robust feature.

% =====================================================================
\subsection{The loops reflect co-firing, not the electrode layout}
\label{sec:r5}

\begin{figure}[t]
\centering
\begin{overpic}[width=\textwidth]{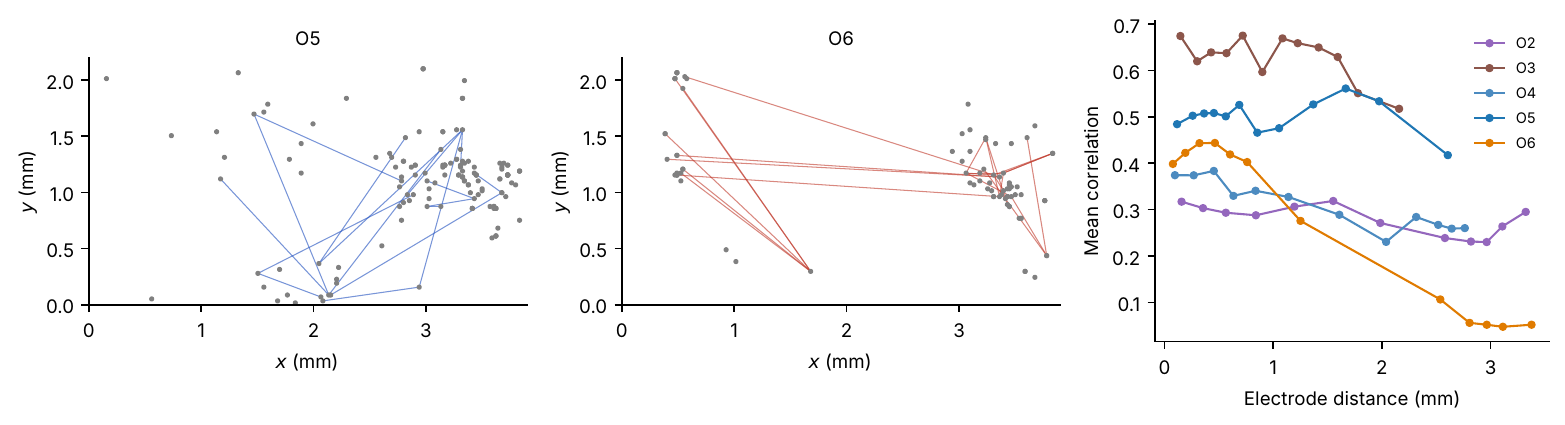}
\put(0.5,27){\panel{a}}\put(34,27){\panel{b}}\put(70,27){\panel{c}}
\end{overpic}
\caption{Where the loops occur. (a,b) Unit positions on the array (mm, $x$ along its
long axis) with the edges of the $H_1$ cocycle supports overlaid, for O5 (a) and O6 (b).
In both, the supports span the array. (c) Mean correlation versus electrode distance for
O2--O6, in bins of equal pair count.}
\label{fig:geometry}
\end{figure}

A correlation network could in principle report the geometry of the recording
rather than the organization of the activity, and for the five datasets that retain
electrode coordinates (O2--O6) we can check. Correlation falls with electrode distance
in four of them (Spearman $\rho=-0.10$ to $-0.71$ over all unit pairs) and not in O5
($0.01$) (Fig.~\ref{fig:geometry}c). If the loops were inherited from the layout, the
units in their $H_1$ cocycle supports would cluster on the array. In O2, O4, O5 and O6
their mean electrode distance is $0.96$ to $1.13$ times that of random unit sets of the
same size, and none is more compact than random sets ($p\ge0.20$), not even O6, whose
correlations depend strongly on distance (Fig.~\ref{fig:geometry}b). In O3, whose
support holds only three units, the ratio is $0.40$ ($p=0.06$).

% =====================================================================
\section{Conclusion}
\label{sec:conclusion}

Persistent homology resolves real, structured loop topology in spontaneous organoid
activity at the modest node counts that recordings provide. The loop structure exceeds a null
that matches firing rate and population bursting, concentrates in a non-redundant
core of carrying units, and is joined by second homology ($H_2$) in the larger
networks. Neither the electrode layout, in the five datasets with coordinates, nor the lag
window, in $13$ of the $14$ significant datasets, accounts for this structure.

Topological analysis of neural systems has been held back by the suspicion that
meaningful Betti numbers require unattainably many nodes. Our results show otherwise.
Across eighteen datasets, $H_1$ structure is resolvable above the null in $9$ of the
$11$ datasets with $N\ge100$ and in five of the seven smaller ones. This
shows persistent homology to be a usable instrument for the recordings the field already
has, and it points to a clear next step: applying the same pipeline to optical
recordings with single-cell resolution and, ultimately, to three-dimensional recording
at high temporal resolution. Because a planar array captures only a fraction of the
units in a three-dimensional culture, raising the unit count this way should bring out
more of the second homology that appears here. A directed filtration would
add the firing order the undirected construction sets aside, and applying the same null
test across the two protocols would show whether the carrying core is conserved across them.

\bigskip
\noindent\textit{Acknowledgements.} We thank Ginestra Bianconi and Michael Freedman for helpful insight into multiple aspects of simplicial complexes and topological data analysis. We acknowledge early experimental support and discussions with Raymond Griffard and Tjitse van der Molen, and theoretical support via discussions with Om Biyani and Bismah Rizwan. We thank
Bismah Rizwan for a close reading of the analysis code.  This work was performed in part with support by the Google Academic Research Award
program (LDC, KSK, DB) and the U.S. National Science Foundation under Grants DGE-2125899 (MB, LDC) and PHY-2515059 (LDC).  NM gratefully acknowledges funding from the Noyce Foundation. LDC thanks the Institute
of Science and Technology Austria for hospitality, and Maksym Serbyn for hosting him
during the sabbatical in which this work was revised and completed.

\bigskip
\noindent\textit{Data availability.} The data and analysis code that support the
findings of this study are available upon request from the authors.

\bigskip
\noindent\textit{CRediT.} Conceptualization (all); Data curation (EB, MB); Formal analysis (EB, MB, LDC); Funding acquisition (LDC, KSK); Investigation (EB, SH, LF); Methodology (EB, MB, NM, LDC); Project administration (LDC, DB, KSK); Resources (KSK); Software (EB, MB, LDC); Supervision (LDC, DB, KSK); Validation (EB, MB, LDC); Visualization (all); Writing – original draft (EB, LDC, DB); Writing – review \& editing (all)

\vspace{18pt}

% =====================================================================
\appendix
\section{Per-dataset statistics and null diagnostics}
\label{app:stats}

This appendix collects the per-dataset results and the checks on the surrogate null.
Table~\ref{tab:stats} lists for each dataset the integrated Betti numbers
$\bar\beta_k=\int\beta_k(\rho)\,d\rho$ of the data, the median and $97.5$th percentile of
the null, $r$ and $q$ (Sec.~\ref{sec:r1}). For the lag-aware duplicate criterion it gives
the unit counts and $r$ out of $19$ surrogates. The last two columns are the removal ratios
of Sec.~\ref{sec:r2}, targeted over random (T/R) and targeted over strength-matched (T/S).

\begin{figure}[h]
\centering
\includegraphics[width=\textwidth]{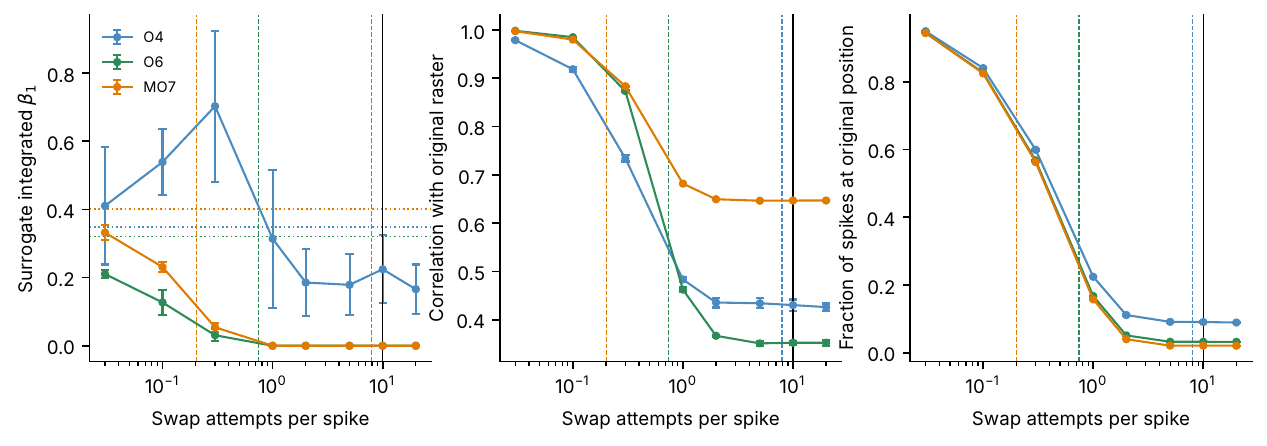}
\caption{Convergence of the raster-marginals chain (mean $\pm$ sd). (Left) Integrated
$\beta_1$ of the surrogates, data dotted. (Centre) Correlation of surrogate with data
matrix. (Right) Fraction of spikes at their original position. Dashed: $10^5$ attempts;
solid: $10$ attempts per spike.}
\label{fig:convergence}
\end{figure}

\begin{figure}[h]
\centering
\includegraphics[width=0.8\textwidth]{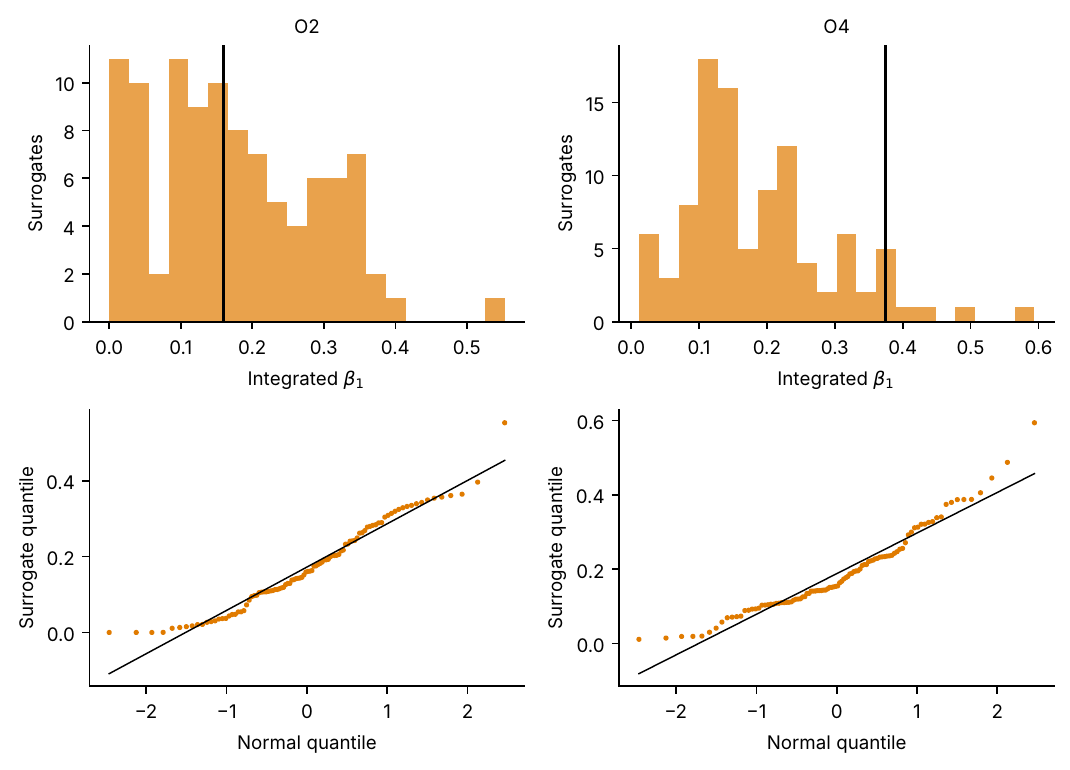}
\caption{Null distributions of integrated $\beta_1$ for O2 and O4 (data as vertical
line), with normal QQ plots below.}
\label{fig:qq}
\end{figure}

\begin{figure}[h]
\centering
\includegraphics[width=0.85\textwidth]{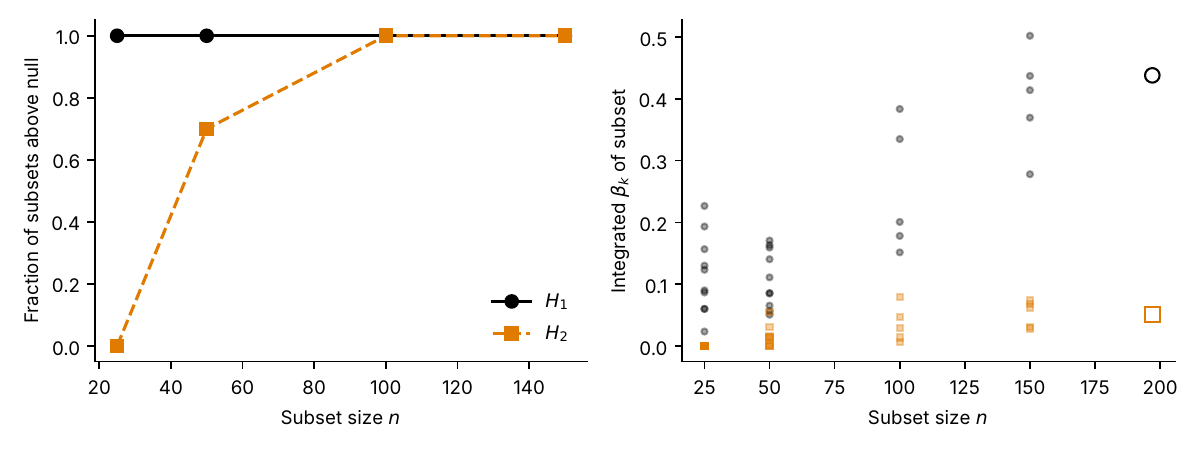}
\caption{Subsampling MO7. (Left) Fraction of subsets that exceed their own null in
$\beta_1$ (black) and $\beta_2$ (orange), against subset size $n$. (Right) Integrated
$\beta_1$ (grey) and $\beta_2$ (orange) of each subset; open symbols, full network.}
\label{fig:subsample}
\end{figure}

Fig.~\ref{fig:convergence}, computed before duplicate removal, shows that the surrogate
chain converges by $2$ attempts per spike, well below the $10$ used here, while a fixed $10^5$
attempts (dashed lines) leaves MO7 and O6 unconverged. Fig.~\ref{fig:qq} shows the null
distributions of O2 and O4, the only datasets above $50$ units whose null is not
concentrated at zero. Both are bounded at zero and skewed. Fig.~\ref{fig:subsample} shows the subsampling of MO7 described in
Sec.~\ref{sec:r3}.

\begin{table}[h]
\centering
\scriptsize
\setlength{\tabcolsep}{3.5pt}
\caption{Per-dataset statistics, ordered by $N$. Null: median [$97.5$th percentile].}
\label{tab:stats}
\begin{tabular}{lrrlrrrrrrrrrr}
\toprule
 & & \multicolumn{4}{c}{$H_1$} & \multicolumn{3}{c}{$H_2$} & \multicolumn{3}{c}{Lag-aware} & \multicolumn{2}{c}{Removal}\\
\cmidrule(lr){3-6}\cmidrule(lr){7-9}\cmidrule(lr){10-12}\cmidrule(lr){13-14}
Dataset & $N$ & $\bar\beta_1$ & null & $r$ & $q$ & $\bar\beta_2$ & $r$ & $q$ & $N$ & $r_1$ & $r_2$ & T/R & T/S\\
\midrule
O1 & 25 & 0 & 0 [0.0051] & 100 & 1.000 & 0 & 100 & 1.000 & 26 & 19 & 19 & -- & -- \\
MO11 & 28 & 0.0026 & 0.0026 [0.024] & 58 & 0.619 & 0 & 100 & 1.000 & 66 & 0 & 19 & -- & -- \\
O3 & 36 & 0.021 & 0 [0.0033] & 1 & 0.025 & 0 & 100 & 1.000 & 40 & 0 & 19 & -- & -- \\
MO12 & 44 & 0.0021 & 0 [0] & 0 & 0.014 & 0 & 100 & 1.000 & 44 & 0 & 19 & -- & -- \\
MO9 & 78 & 0.013 & 0 [0] & 0 & 0.014 & 0 & 100 & 1.000 & 81 & 0 & 19 & 1.66 & 1.46 \\
O6 & 85 & 0.113 & 0 [0] & 0 & 0.014 & 0.013 & 0 & 0.020 & 85 & 0 & 0 & 1.56 & 2.55 \\
MO4 & 93 & 0.207 & 0 [0.0002] & 0 & 0.014 & 0 & 100 & 1.000 & 99 & 0 & 19 & 1.40 & 3.17 \\
MO3 & 113 & 0.152 & 0 [0] & 0 & 0.014 & 0.0054 & 0 & 0.020 & 124 & 0 & 0 & 1.31 & 2.39 \\
O2 & 120 & 0.160 & 0.160 [0.363] & 50 & 0.568 & 0 & 100 & 1.000 & 130 & 10 & 19 & 1.55 & 1.99 \\
O4 & 122 & 0.374 & 0.154 [0.426] & 8 & 0.107 & 0 & 100 & 1.000 & 123 & 2 & 19 & 1.53 & 3.49 \\
MO10 & 125 & 0.356 & 0 [0] & 0 & 0.014 & 0.0027 & 0 & 0.020 & 136 & 0 & 0 & 1.48 & 3.49 \\
O5 & 146 & 0.045 & 0 [0] & 0 & 0.014 & 0.0019 & 0 & 0.020 & 148 & 0 & 0 & 1.48 & 1.27 \\
MO6 & 146 & 0.138 & 0 [0] & 0 & 0.014 & 0.0014 & 0 & 0.020 & 148 & 0 & 0 & 1.63 & 3.68 \\
MO2 & 165 & 0.200 & 0 [0.0001] & 0 & 0.014 & 0 & 100 & 1.000 & 179 & 0 & 19 & 1.55 & 2.88 \\
MO8 & 168 & 0.117 & 0 [0] & 0 & 0.014 & 0.032 & 0 & 0.020 & 171 & 0 & 0 & 1.59 & 1.07 \\
MO1 & 179 & 0.164 & 0 [0] & 0 & 0.014 & 0.0013 & 0 & 0.020 & 197 & 0 & 0 & 1.55 & 1.68 \\
MO5 & 188 & 0.314 & 0 [0] & 0 & 0.014 & 0.014 & 0 & 0.020 & 189 & 0 & 0 & 1.23 & 2.50 \\
MO7 & 197 & 0.438 & 0 [0] & 0 & 0.014 & 0.051 & 0 & 0.020 & 205 & 0 & 0 & 1.21 & 3.10 \\
\bottomrule
\end{tabular}
\end{table}

\clearpage
\bibliographystyle{iopart-num}
\bibliography{references}

@article{Sharf2022Functional,
  title={Functional neuronal circuitry and oscillatory dynamics in human brain organoids},
  author={Sharf, T. and van der Molen, T. and Glasauer, S.M.K. and others},
  journal={Nature Communications},
  volume={13},
  pages={4403},
  year={2022},
  doi={10.1038/s41467-022-32115-4},
  url={https://doi.org/10.1038/s41467-022-32115-4}
}

@article{Singh2008,
  author    = {Gurjeet Singh and Facundo Memoli and Tigran Ishkhanov and Guillermo Sapiro and Gunnar Carlsson and Dario L. Ringach},
  title     = {Topological analysis of population activity in visual cortex},
  journal   = {Journal of Vision},
  volume    = {8},
  number    = {8},
  pages     = {11},
  year      = {2008},
  month     = {June},
  doi       = {10.1167/8.8.11},
  url       = {https://doi.org/10.1167/8.8.11},
  publisher = {ARVO Journals},
}

@article{gardner_2022,
  author    = {Gardner, R. J. and Hermansen, E. and Pachitariu, M. and others},
  title     = {Toroidal topology of population activity in grid cells},
  journal   = {Nature},
  volume    = {602},
  pages     = {123--128},
  year      = {2022},
  doi       = {10.1038/s41586-021-04268-7},
}

@inproceedings{Acosta2022QuantifyingLE,
  title={Quantifying Extrinsic Curvature in Neural Manifolds},
  author={Acosta, Francisco and Sanborn, Sophia and Dao Duc, Khanh and Madhav, Manu and Miolane, Nina},
  booktitle={Proceedings of the IEEE/CVF Conference on Computer Vision and Pattern Recognition (CVPR) Workshops},
  pages={610--619},
  year={2023}
}

@article{Pachitariu2024,
  author    = {Marius Pachitariu and Shashwat Sridhar and Jacob Pennington and Carsen Stringer},
  title     = {Spike sorting with Kilosort4},
  journal   = {Nature Methods},
  volume    = {21},
  number    = {5},
  pages     = {914--921},
  year      = {2024},
  doi       = {10.1038/s41592-024-02232-7},
  url       = {https://doi.org/10.1038/s41592-024-02232-7},
  publisher = {Nature Publishing Group},
}

@article{Birey2017,
  author    = {Fikri Birey and Jimena Andersen and Christopher D. Makinson and Saiful Islam and Wu Wei and Nina Huber and H. Christina Fan and Kimberly R. Cordes Metzler and Georgia Panagiotakos and Nicholas Thom and Nancy A. O’Rourke and Lars M. Steinmetz and Jonathan A. Bernstein and Joachim Hallmayer and John R. Huguenard and Sergiu P. Paşca},
  title     = {Assembly of functionally integrated human forebrain spheroids},
  journal   = {Nature},
  volume    = {545},
  pages     = {54--59},
  year      = {2017},
  month     = {April},
  doi       = {10.1038/nature22330},
  url       = {https://www.nature.com/articles/nature22330},
  publisher = {Nature Publishing Group},
}

@article{Lancaster2013,
  author    = {Madeline A. Lancaster and Magdalena Renner and Carol-Anne Martin and Daniel Wenzel and Louise S. Bicknell and Matthew E. Hurles and Tessa Homfray and Josef M. Penninger and Andrew P. Jackson and Juergen A. Knoblich},
  title     = {Cerebral organoids model human brain development and microcephaly},
  journal   = {Nature},
  volume    = {501},
  number    = {7467},
  pages     = {373--379},
  year      = {2013},
  month     = {September},
  doi       = {10.1038/nature12517},
  url       = {https://doi.org/10.1038/nature12517},
  publisher = {Nature Publishing Group},
}

@article{ctralie2018ripser,
    doi = {10.21105/joss.00925},
    url = {https://doi.org/10.21105/joss.00925},
    year  = {2018},
    month = {Sep},
    publisher = {The Open Journal},
    volume = {3},
    number = {29},
    pages = {925},
    author = {Christopher Tralie and Nathaniel Saul and Rann Bar-On},
    title = {{Ripser.py}: A Lean Persistent Homology Library for Python},
    journal = {The Journal of Open Source Software}
}

@article{Giusti2015,
  author  = {Giusti, Chad and Pastalkova, Eva and Curto, Carina and Itskov, Vladimir},
  title   = {Clique topology reveals intrinsic geometric structure in neural correlations},
  journal = {Proceedings of the National Academy of Sciences},
  volume  = {112},
  number  = {44},
  pages   = {13455--13460},
  year    = {2015},
  doi     = {10.1073/pnas.1506407112},
  url     = {https://doi.org/10.1073/pnas.1506407112}
}

@article{Sizemore2018,
  author  = {Sizemore, Ann E. and Giusti, Chad and Kahn, Ari and Vettel, Jean M. and Betzel, Richard F. and Bassett, Danielle S.},
  title   = {Cliques and cavities in the human connectome},
  journal = {Journal of Computational Neuroscience},
  volume  = {44},
  number  = {1},
  pages   = {115--145},
  year    = {2018},
  doi     = {10.1007/s10827-017-0672-6},
  url     = {https://doi.org/10.1007/s10827-017-0672-6}
}

@article{Petri2014,
  author  = {Petri, G. and Expert, P. and Turkheimer, F. and Carhart-Harris, R. and Nutt, D. and Hellyer, P. J. and Vaccarino, F.},
  title   = {Homological scaffolds of brain functional networks},
  journal = {Journal of the Royal Society Interface},
  volume  = {11},
  number  = {101},
  pages   = {20140873},
  year    = {2014},
  doi     = {10.1098/rsif.2014.0873},
  url     = {https://doi.org/10.1098/rsif.2014.0873}
}

@article{Sizemore2019,
  author  = {Sizemore, Ann E. and Phillips-Cremins, Jennifer E. and Ghrist, Robert and Bassett, Danielle S.},
  title   = {The importance of the whole: Topological data analysis for the network neuroscientist},
  journal = {Network Neuroscience},
  volume  = {3},
  number  = {3},
  pages   = {656--673},
  year    = {2019},
  doi     = {10.1162/netn_a_00073},
  url     = {https://doi.org/10.1162/netn_a_00073}
}

@article{Reimann2017,
  author  = {Reimann, Michael W. and Nolte, Max and Scolamiero, Martina and Turner, Katharine and Perin, Rodrigo and Chindemi, Giuseppe and D{\l}otko, Pawe{\l} and Levi, Ran and Hess, Kathryn and Markram, Henry},
  title   = {Cliques of neurons bound into cavities provide a missing link between structure and function},
  journal = {Frontiers in Computational Neuroscience},
  volume  = {11},
  pages   = {48},
  year    = {2017},
  doi     = {10.3389/fncom.2017.00048},
  url     = {https://doi.org/10.3389/fncom.2017.00048}
}

@article{MEANAP2024,
  author  = {Sit, Timothy P. H. and Feord, Rachael C. and Dunn, Alexander W. E. and others},
  title   = {{MEA-NAP}: A flexible network analysis pipeline for neuronal 2{D} and 3{D} organoid multielectrode recordings},
  journal = {Cell Reports Methods},
  volume  = {4},
  number  = {11},
  pages   = {100901},
  year    = {2024},
  doi     = {10.1016/j.crmeth.2024.100901}
}

@article{Okun2012,
  author  = {Okun, Michael and Yger, Pierre and Marguet, Stephan L. and Gerard-Mercier, Florian and Benucci, Andrea and Katzner, Steffen and Busse, Laura and Carandini, Matteo and Harris, Kenneth D.},
  title   = {Population Rate Dynamics and Multineuron Firing Patterns in Sensory Cortex},
  journal = {Journal of Neuroscience},
  volume  = {32},
  number  = {48},
  pages   = {17108--17119},
  year    = {2012},
  doi     = {10.1523/JNEUROSCI.1831-12.2012}
}

@article{Ryser1957,
  author  = {Ryser, H. J.},
  title   = {Combinatorial properties of matrices of zeros and ones},
  journal = {Canadian Journal of Mathematics},
  volume  = {9},
  pages   = {371--377},
  year    = {1957},
  doi     = {10.4153/CJM-1957-044-3}
}

@article{Battiston2020,
  author  = {Battiston, Federico and Cencetti, Giulia and Iacopini, Iacopo and Latora, Vito and Lucas, Maxime and Patania, Alice and Young, Jean-Gabriel and Petri, Giovanni},
  title   = {Networks beyond pairwise interactions: Structure and dynamics},
  journal = {Physics Reports},
  volume  = {874},
  pages   = {1--92},
  year    = {2020},
  doi     = {10.1016/j.physrep.2020.05.004}
}

@article{Battiston2021,
  author  = {Battiston, Federico and Amico, Enrico and Barrat, Alain and Bianconi, Ginestra and Ferraz de Arruda, Guilherme and Franceschiello, Benedetta and Iacopini, Iacopo and K{\'e}fi, Sonia and Latora, Vito and Moreno, Yamir and Murray, Micah M. and Peixoto, Tiago P. and Vaccarino, Francesco and Petri, Giovanni},
  title   = {The physics of higher-order interactions in complex systems},
  journal = {Nature Physics},
  volume  = {17},
  number  = {10},
  pages   = {1093--1098},
  year    = {2021},
  doi     = {10.1038/s41567-021-01371-4}
}

@article{BassettSporns2017,
  author  = {Bassett, Danielle S. and Sporns, Olaf},
  title   = {Network neuroscience},
  journal = {Nature Neuroscience},
  volume  = {20},
  number  = {3},
  pages   = {353--364},
  year    = {2017},
  doi     = {10.1038/nn.4502}
}

@article{Schreiber2003,
  author  = {Schreiber, Susanne and Fellous, Jean-Marc and Whitmer, Diane and Tiesinga, Paul and Sejnowski, Terrence J.},
  title   = {A New Correlation-Based Measure of Spike Timing Reliability},
  journal = {Neurocomputing},
  volume  = {52--54},
  pages   = {925--931},
  year    = {2003},
  doi     = {10.1016/S0925-2312(02)00838-X}
}

@article{Osaki2024,
  author  = {Osaki, Tatsuya and Duenki, Tomoya and Chow, Siu Yu A. and others},
  title   = {Complex activity and short-term plasticity of human cerebral organoids reciprocally connected with axons},
  journal = {Nature Communications},
  volume  = {15},
  pages   = {2945},
  year    = {2024},
  doi     = {10.1038/s41467-024-46787-7}
}

@article{Wilson2022,
  author  = {Wilson, Madison N. and Thunemann, Martin and Liu, Xin and others},
  title   = {Multimodal monitoring of human cortical organoids implanted in mice reveal functional connection with visual cortex},
  journal = {Nature Communications},
  volume  = {13},
  pages   = {7945},
  year    = {2022},
  doi     = {10.1038/s41467-022-35536-3}
}

@article{Adamaszek2017,
  author  = {Adamaszek, Micha{\l} and Adams, Henry},
  title   = {The {V}ietoris--{R}ips complexes of a circle},
  journal = {Pacific J. Math.},
  volume  = {290},
  number  = {1},
  pages   = {1--40},
  year    = {2017},
  doi     = {10.2140/pjm.2017.290.1}
}

@article{Benjamini1995,
  author  = {Benjamini, Yoav and Hochberg, Yosef},
  title   = {Controlling the false discovery rate: a practical and powerful approach to multiple testing},
  journal = {J. R. Stat. Soc. B},
  volume  = {57},
  number  = {1},
  pages   = {289--300},
  year    = {1995},
  doi     = {10.1111/j.2517-6161.1995.tb02031.x}
}

@article{Phipson2010,
  author  = {Phipson, Belinda and Smyth, Gordon K.},
  title   = {Permutation {P}-values should never be zero: calculating exact {P}-values when permutations are randomly drawn},
  journal = {Stat. Appl. Genet. Mol. Biol.},
  volume  = {9},
  number  = {1},
  pages   = {39},
  year    = {2010},
  doi     = {10.2202/1544-6115.1585}
}

@article{Miklos2004,
  author  = {Mikl{\'o}s, Istv{\'a}n and Podani, J{\'a}nos},
  title   = {Randomization of presence--absence matrices: comments and new algorithms},
  journal = {Ecology},
  volume  = {85},
  number  = {1},
  pages   = {86--92},
  year    = {2004},
  doi     = {10.1890/03-0101}
}

@article{Okun2015,
  author  = {Okun, Michael and Steinmetz, Nicholas A. and Cossell, Lee and Iacaruso, M. Florencia and Ko, Ho and Barth{\'o}, P{\'e}ter and Moore, Tirin and Hofer, Sonja B. and Mrsic-Flogel, Thomas D. and Carandini, Matteo and Harris, Kenneth D.},
  title   = {Diverse coupling of neurons to populations in sensory cortex},
  journal = {Nature},
  volume  = {521},
  pages   = {511--515},
  year    = {2015},
  doi     = {10.1038/nature14273}
}

\end{document}